\documentclass[aps,prl,twocolumn,superscriptaddress,showpacs]{revtex4-1}

\usepackage{graphicx}
\usepackage{color}

\begin{document}


\title{The first direct measurement of {}$^{12}$C({}$^{12}$C,n){}$^{23}$Mg at
 stellar energies}


\author{B. Bucher}
\email[]{bucher3@llnl.gov}
\affiliation{Institute for Structure and Nuclear Astrophysics, Joint
 Institute for Nuclear Astrophysics, University of Notre Dame, Notre Dame,
 Indiana 46556, USA}
\affiliation{Lawrence Livermore National Laboratory, Livermore, California
 94550, USA}
\author{X. D. Tang}
\email[]{xtang@impcas.ac.cn}
\affiliation{Institute of Modern Physics, Chinese Academy of Science,
 Lanzhou, Gansu 730000, P.R. China}
\author{X. Fang}
\affiliation{Institute for Structure and Nuclear Astrophysics, Joint
 Institute for Nuclear Astrophysics, University of Notre Dame, Notre Dame,
 Indiana 46556, USA}
\author{A. Heger}
\altaffiliation{NuGrid collaboration, \url{http://www.nugridstars.org}}
\affiliation{Monash Center for Astrophysics, School of Mathematical Sciences,
 Monash University, Victoria 3800, Australia}
\author{S. Almaraz-Calderon}
\altaffiliation{Present address: Department of Physics, Florida State
 University, Tallahassee, Florida 32306, USA}
\author{A. Alongi} 
\author{A. D. Ayangeakaa}
\altaffiliation{Present address: Physics Division, Argonne National
 Laboratory, Argonne, Illinois 60439 USA}
\author{M. Beard}
\author{A. Best}
\altaffiliation{Present address: INFN, Laboratori Nazionali del Gran Sasso
 (LNGS), 67010 Assergi, Italy}
\author{J. Browne}
\author{C. Cahillane}
\author{M. Couder}
\author{R. J. deBoer}
\author{A. Kontos}
\affiliation{Institute for Structure and Nuclear Astrophysics, Joint
 Institute for Nuclear Astrophysics, University of Notre Dame, Notre Dame,
 Indiana 46556, USA}
\author{L. Lamm}
\altaffiliation{Deceased}
\affiliation{Institute for Structure and Nuclear Astrophysics, Joint
 Institute for Nuclear Astrophysics, University of Notre Dame, Notre Dame,
 Indiana 46556, USA}
\author{Y. J.~Li}
\affiliation{China Institute of Atomic Energy, Beijing 102413, P.R. China}
\author{A. Long}
\author{W. Lu}
\author{S. Lyons}
\author{M. Notani}
\author{D. Patel}
\author{N. Paul}
\affiliation{Institute for Structure and Nuclear Astrophysics, Joint
 Institute for Nuclear Astrophysics, University of Notre Dame, Notre Dame,
 Indiana 46556, USA}
\author{M. Pignatari}
\altaffiliation{NuGrid collaboration, \url{http://www.nugridstars.org}}
\affiliation{Konkoly Observatory, Research Centre for Astronomy and Earth
 Sciences,
Hungarian Academy of Sciences, Konkoly Thege Mikl\'os \'ut 15-17, H-1121
 Budapest, Hungary}
\affiliation{Department of Physics, University of Basel, Basel, CH-4056,
 Switzerland}
\author{A. Roberts}
\author{D. Robertson}
\author{K. Smith}
\author{E. Stech}
\author{R. Talwar}
\author{W. P. Tan}
\author{M. Wiescher}
\affiliation{Institute for Structure and Nuclear Astrophysics, Joint
 Institute for Nuclear Astrophysics, University of Notre Dame, Notre Dame,
 Indiana 46556, USA}
\author{S. E. Woosley}
\affiliation{Department of Astronomy and Astrophysics, University of
 California, Santa Cruz, California 95064, USA}


\date{\today}

\begin{abstract}
Neutrons produced by the carbon fusion reaction
${}^{12}$C(${}^{12}$C,n)${}^{23}$Mg
play an important role in
stellar nucleosynthesis.  However, past studies have shown 
large discrepancies between experimental data and theory, leading to an
 uncertain
cross section extrapolation at astrophysical energies.
We present the first direct
measurement that extends deep into the
astrophysical energy range along with a
new and improved extrapolation technique based on experimental data
from the mirror reaction 
${}^{12}$C(${}^{12}$C,p)${}^{23}$Na.
The new reaction rate has been determined with a well-defined uncertainty
 that exceeds
the precision required by astrophysics models.
Using our constrained rate, we find that ${}^{12}$C(${}^{12}$C,n)${}^{23}$Mg 
is crucial to the production of Na and Al in Pop-III Pair Instability
 Supernovae.
It also plays a non-negligible role in the production of weak s-process
 elements
 as well as in the production of the important galactic $\gamma$-emitter 
${}^{60}$Fe.
\end{abstract}


\maketitle


The first stars in the early Universe formed about 400 million years
after the big bang.  Verification of the existence of these stars is
important for our understanding of the evolution of the Universe 
\cite{Bromm2013}. 
It has been predicted that for Population-III (metal-free stars 
\cite{LeBlanc2010}) stellar production yields, 
the abundances of odd-Z elements are remarkably deficient compared to their
 adjacent even-Z elements \cite{heger2002}. 
Astronomers are searching for long-lived, low mass stars with the
unique nucleosynthetic pattern matching the predicted yields
\cite{AOKI2014}. The relevance of ${}^{12}$C(${}^{12}$C,n)${}^{23}$Mg in the
 first stars has been discussed by Woosley, Heger, and Weaver \cite{whw2002}.
By the end of helium burning in Pop-III stars, the neutron to proton ratio in
 the ash is almost exactly 1. 
However, in the subsequent carbon burning phase, frequent $\beta^{+}$ decay
 of produced ${}^{23}$Mg 
converts protons into neutrons, thus increasing the neutron to proton ratio. 
A slight excess of neutrons would significantly affect the
abundances of the odd-Z isotopes with neutron to proton ratios higher than 1,
e.g. ${}^{23}$Na and ${}^{27}$Al. 

${}^{12}$C(${}^{12}$C,n)$^{23}$Mg is also a potentially important
neutron source for the so-called weak s-process occurring 
in massive Pop-I (metal-rich \cite{LeBlanc2010}) and Pop-II 
(metal-poor \cite{LeBlanc2010}) stars.  The weak s-process takes 
place during the core helium and shell carbon burning phases and 
is largely responsible for the s-process abundances up to A$\approx$90
\cite{Pignatari2010}. 
Pignatari et
al. recently performed a study of the weak s-process during carbon shell
burning for a 25\,M$_\odot$ stellar model using different
${}^{12}$C(${}^{12}$C,n)$^{23}$Mg rates \cite{Pignatari2013}. They
found that a factor of 2 precision or better would be desirable to
limit its impact on the s-process predictions to within 10\%.

\begin{figure}
    \includegraphics[width=20pc]{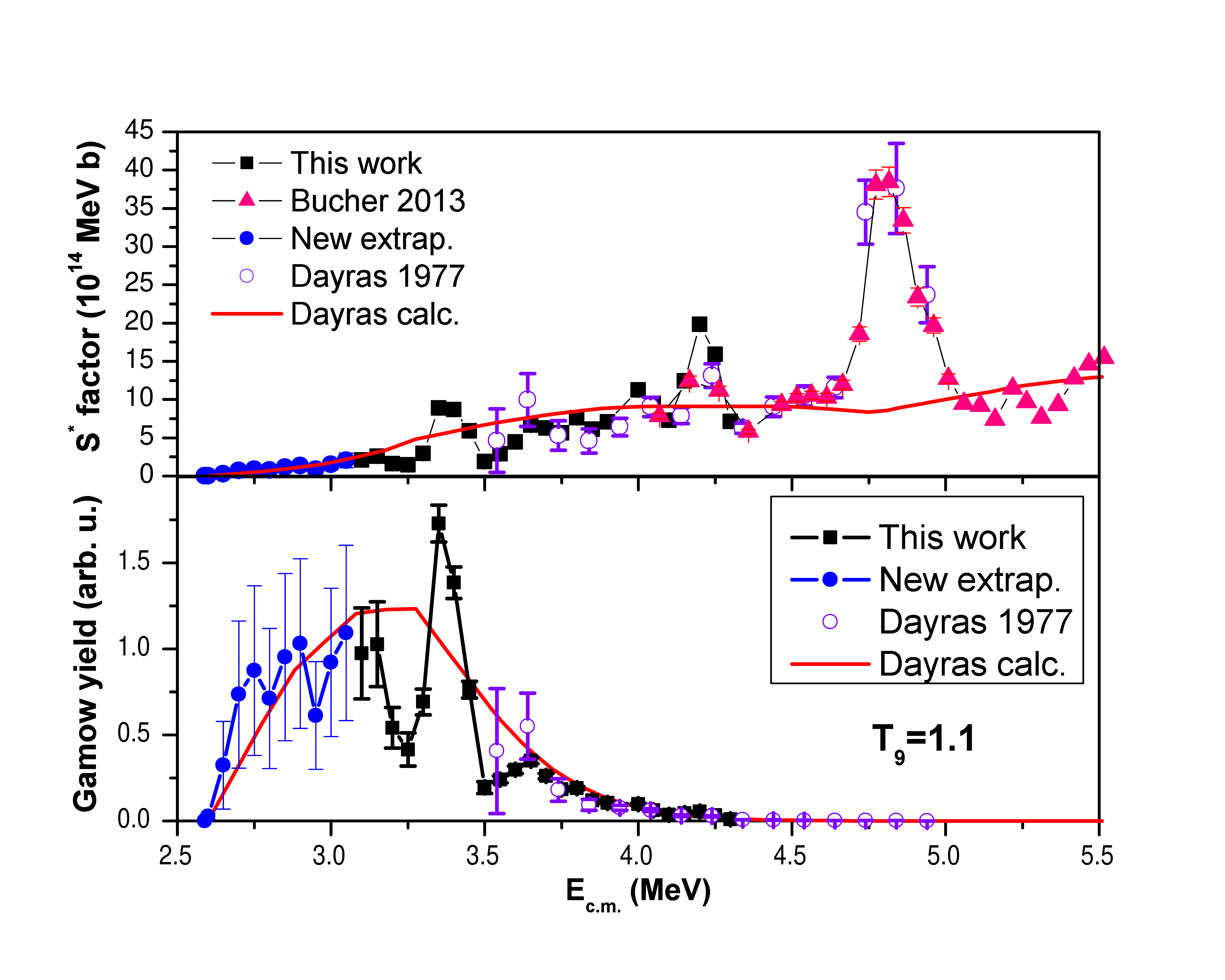}
    \caption{(Color online) Upper: The $^{12}$C($^{12}$C,n)$^{23}$Mg 
S*-factor results from the present measurement (black squares) compared with
 previous data sets from Dayras~1977 \cite{Dayras1977} 
(purple circles) and
 Bucher~2013 \cite{bucherThesis,bucherNN2012} 
(magenta triangles).  Also
 shown is the neutron branching ratio calculated by Dayras applied to the
 total $^{12}$C+$^{12}$C fusion S*-factor recommended by CF88 
\cite{Caughlan1988} (red solid line) and the new extrapolation from this work
 (blue circles).  Only statistical errors are shown for the experimental
 data, while the extrapolation includes both statistical and a 40\%
 systematic error. Lower: The integrand from Eq.~\ref{eq:reaction_rate} using
 the various data sets is plotted.  }
    \label{fig:sFactor}
\end{figure}

Stellar carbon burning has three main reaction channels:
\begin{eqnarray*}
{}^{12}\rm C+{}^{12}\rm C &\rightarrow& {}^{23}\rm Mg+{\rm n}-2.60\text{ MeV}
 \nonumber
\\                          &\rightarrow& {}^{23}{\rm Na}+{\rm p}+2.24
\text{ MeV} \nonumber
\\                          &\rightarrow& {}^{20}{\rm Ne}+\alpha+4.62
\text{ MeV}
\end{eqnarray*}
With $Q<0$, the probability of decay through the neutron channel is weakest
 among
the three at the low energies relevant for astrophysics. 
For a typical carbon shell
burning temperature T$_9=1.1$, the important energy range for this
channel is  2.7$\,<\,$E$_{\mathrm{cm}}$$\,<\,$3.6 MeV.  
The reaction was first studied in 1969 by Patterson et al. 
\cite{Patterson1969}
who measured
the cross section over the range E$_{\mathrm{cm}}$=4.23 to 8.74 MeV by
 counting
$\beta$-rays from $^{23}$Mg decays.
From this measurement, a constant neutron branching ratio, $\beta_n$=\,2\%,
 was deduced \cite{at69}. 
Later Dayras et al. extended the measurement down to
E$_{\mathrm{cm}}$=3.54 MeV by counting the $\gamma$-rays emitted
following the $^{23}$Mg beta decay.  The experimental uncertainty is
about 40\% at E$_{\mathrm{cm}}\approx3.8\,$MeV and increases
to 90\% at the lowest energy \cite{Dayras1977}.  To estimate the
cross section at the stellar burning energies, Dayras et al. had to rely on
an extrapolation of the experimental data based on a Hauser-Feshbach
statistical model calculation \cite{Dayras1976}.  Because of the unique
 molecular resonances
existing in the $^{12}$C+$^{12}$C fusion reaction \cite{AKB60}, their
 calculation could only 
be renormalized to the average trend of the data while 
the resonant behavior of the $^{12}$C+$^{12}$C fusion
reaction was ignored.  The maximum deviation between the experimental result
 and
the renormalized statistical model prediction is more than a factor of
$4$ (see Fig.~\ref{fig:sFactor}).  
Nevertheless, based on the statistical model extrapolation, this
work recommended a neutron branching ratio of ${\beta}_n$=\,0.011\%,
0.11\%, 0.40\% and 5.4\% at $T_9$=\,0.8, 1.0, 1.2, and 5, respectively,
though no attempt was made to quantify the uncertainties in these predictions
 \cite{Dayras1977}.  

In 1988, Caughlan and Fowler (CF88) excluded this result from their rate
compilations \cite{Caughlan1988}.  Instead, they recommended
$\beta_n$=\,0 ($T_9$$<\,$1.75), $\beta_n$=\,5\% (1.75$\,\leq$\,$T_9$$<\,$3.3)
 and
$\beta_n$=\,7\% (3.3$\,\leq\,$$T_9$$\,<\,$6.0).  This rate was adopted by
REACLIB after fitting the CF88 ratio with the standard REACLIB
formula \cite{reaclib}.
Pignatari et al. attempted to use the Dayras rate \cite{Pignatari2013},
 however it was later discovered that the analytic formula for ${\beta}_n$
 taken from the paper \cite{Dayras1977} contained a typographical error resulting in a significant deviation from the intended value below T$_9$
=\,1.5 \cite{bucherNN2012}.  
So far, to our knowledge, the correct Dayras rate has only been implemented
 in the stellar code KEPLER \cite{Kepler1,Kepler2}.  

In the following, we report on the first direct measurement of this 
reaction into the stellar energy range as well as an improved method for
extrapolating the experimental results through the remaining unmeasured
 energies relevant
for carbon shell burning. Based on the
new experimental result, a new reaction rate is recommended together
with a well-defined uncertainty. The impact on the nucleosynthesis in
massive stars is also discussed.

The experimental work was performed at the University of Notre Dame's Nuclear
 Science Laboratory using the 11 MV FN tandem Van de Graaff accelerator.
Carbon beams were produced at energies ranging from 5.1 to 8.7 MeV (lab
 frame) with typical currents on target between 0.5 and 1.5 p$\mu$A.  The
beam energy calibration was checked by measuring the reaction thresholds of 
$^7$Li(p,n) and $^{19}$F(p,n) 
as well as $^{12}$C(p,p) resonant scattering \cite{Wilkerson1983}.  
The maximum energy deviation was less than 0.1\%.
A 1-mm thick hydrogen-free Highly Ordered Pyrolytic Graphite (HOPG) target
 made from natural carbon was used to control the hydrogen-induced background
 \cite{ZickProc2010}.   
The target was cooled by circulating deionized water through the supporting
 flange, which 
was centered in a block of polyethylene containing 20 $^3$He proportional
 counters 
arranged around the beam axis in two concentric rings 
\cite{bucherNN2012,Falahat2013}.

 The main sources of beam-induced neutron background were from the reactions
 $^{13}$C($^{12}$C,n)$^{24}$Mg and, to a lesser extent, 
$^2$H($^{12}$C,$^{13}$N)n \cite{bucherNN2012}. 
With a large positive Q-value (8.99 MeV) and the relatively high natural
 abundance of $^{13}$C in the target (1.1\%), neutrons from 
$^{13}$C($^{12}$C,n)$^{24}$Mg dominate the total yield at very low beam
 energies approaching the $^{12}$C($^{12}$C,n)$^{23}$Mg reaction threshold.  
To estimate its contribution, the $^{13}$C($^{12}$C,n)$^{24}$Mg reaction was
 studied with the same setup using a $^{13}$C beam with energies ranging
 between 9.5 and 5.4 MeV.  Since the cross section for this reaction is much
 higher, relatively low beam intensities ($\simeq$50 pnA) with shorter run
 times were sufficient.  
The normalized $^{12}$C($^{13}$C,n)$^{24}$Mg yield was then subtracted from
 the measured total neutron yield recorded with the $^{12}$C beam
 \cite{bucherThesis}.  

The room background rate was measured to be  
9.015(92) evts/min, which dominated the yield at energies below 
E$_{\mathrm{cm}}$=\,3.0 MeV. The background contribution from 
$^2$H($^{12}$C,$^{13}$N)n was studied using a thin TiD$_2$ target with thick
 Cu backing. After removing the room background, this contribution was found
 to be less than 5\% of the total yield at E$_{\mathrm{cm}}$=\,3.3 MeV
 increasing to 19\% at 3.1 MeV. 

The detector efficiency has been simulated using Geant4 and 
MCNP in the range 0 to 3 MeV and experimentally validated with 
monoenergetic neutrons between 50 and 650 keV \cite{Falahat2013}.  
For this experiment, the Geant4 code was modified to include the strong 
angular dependence of the neutron energy from the 
$^{12}$C($^{12}$C,n)$^{23}$Mg kinematics.  
An isotropic angular distribution in the 
center-of-mass frame was assumed for the neutrons produced by 
${}^{12}$C(${}^{12}$C,n)$^{23}$Mg and ${}^{13}$C(${}^{12}$C,n)$^{24}$Mg. 
The efficiency was found to vary smoothly between 30\% and 50\% as the
 average neutron energy decreases with beam energy \cite{bucherThesis}. 
To check the effect of the assumed angular
 distribution, we changed the isotropic angular distribution to match 
2$^{\rm nd}$ and 4$^{\rm th}$ order Legendre polynomials \cite{Becker1981},
 and found a nearly constant relative drop in efficiency of 9\% and 5\%,
 respectively, in the range of E$_{\rm cm}$=\,3 to 5 MeV. Since our
 experiment does not measure angular distribution, 
a $\pm$5\% systematic uncertainty has been assigned for this effect.
To test our efficiency calculation, we measured the 
${}^{12}$C(${}^{12}$C,n)$^{23}$Mg cross section independently by 
detecting the activity of ${}^{23}$Mg \cite{bucherThesis,bucherNN2012}.  
The results gave good agreement in the overlapping energy range 
(as shown in Fig.~\ref{fig:sFactor}).

The cross section for the ${}^{12}$C(${}^{12}$C,n)$^{23}$Mg reaction 
was determined by differentiating the thick target yield 
\cite{notani2012}.  In Fig.~\ref{fig:sFactor}, it has been converted to 
a modified S-factor (S*) for comparison with previous 
results.  S* differs from the standard S-factor only by the multiplier, 
$\exp(0.46\,E)$, which is added to account for the finite size of the 
nucleus \cite{Patterson1969}.  It is seen that the new results display 
good agreement with previous measurements in the overlapping energy 
region while extending much deeper into the  
astrophysical energy range.  A new resonance at E$_{\rm cm}$=\,3.4 MeV 
is observed in the neutron channel.  This resonance was also 
observed in earlier measurements of the proton and alpha channels 
\cite{Kettner1980,Becker1981,Spillane2007,zickefoose,Fang2013}.   
Our measurement includes a 15\% systematic uncertainty which primarily 
results from the uncertainties in the beam current (10\%), beam energy (2\%), 
detector efficiency (6\%) \cite{Falahat2013}, angular distribution (5\%), 
and stopping power (7\%) \cite{SRIM,bucherThesis}.  The Dayras results 
also have an additional systematic uncertainty of 16\% \cite{Dayras1977} 
not shown in Fig.~\ref{fig:sFactor}.

An extrapolation is required to estimate the reaction cross section at 
the lower energies beyond experimental reach.  As mentioned earlier, 
Dayras et al. provided a renormalized statistical model calculation for 
this purpose.  
However, 
the large discrepancy between the experimental data and their theory
calls into question the reliability of the extrapolation.    
To provide a better prediction including the effect of the 
molecular resonances in the entrance channel, 
a novel extrapolation method has been developed based on 
experimental information from the mirror reaction 
$^{12}$C($^{12}$C,p)$^{23}$Na. The predicted neutron cross section, 
$\sigma_{n(\text{pred})}$, is obtained using the formula
\begin{equation}\label{eq:predict_n}
\sigma_{n(\text{pred})}=
\sum_{i=0}^{N}\frac{\sigma_{\mathrm n_i(\text{th})}}{\sigma_{\mathrm p_i(
\text{th})}}\sigma_{\mathrm p_i(\text{exp})}
\end{equation}
where $N$ is the highest available decay channel in the residual 
$^{23}$Mg, which depends on the reaction energy.  
For E$_{\rm cm}$$\leq$\,4.6 MeV, only the n$_0$ and n$_1$ channels are open. 
The theoretical ratio, $\sigma_{\mathrm n_i(\text{th})}/\sigma_{
\mathrm p_i(\text{th})}$, is calculated using TALYS \cite{TALYS} 
combined with entrance channel spin populations supplied from a 
coupled-channels calculation by Esbensen \cite{Esbensen2011}.
The resonances in ${}^{12}$C(${}^{12}$C,n$_i$)${}^{23}$Mg and 
${}^{12}$C(${}^{12}$C,p$_i$)${}^{23}$Na originate from both the 
molecular resonances in the entrance channel and the 
characteristic resonances in the final decay channels.
The traditional statistical model calculation employed by Dayras uses 
the optical model and assumes a high level density to describe the 
entrance and exit channels and therefore could only reproduce the 
average trend of the experimental data. 
 In our approach, the complicated molecular resonance associated with 
the entrance channel is embodied in the experimental cross sections 
($\sigma_{\mathrm p_i(\text{exp})}$) of 
${}^{12}$C(${}^{12}$C,p$_i$)$^{23}$Na, the mirror system of 
${}^{12}$C(${}^{12}$C,n$_i$)$^{23}$Mg, while the statistical model is 
only used to predict the decay width ratio between the n$_i$ and 
p$_i$ channels. 
Since the proton energy resolution in the Zickefoose experiment from 
Ref.~\cite{zickefoose}
was insufficient to resolve p$_0$ from p$_1$,
only the sum, $\sigma_{\mathrm p_0}+\sigma_{\rm p_1}$, is available for 
E$_{\mathrm{cm}}$$<$\,4 MeV.
Eq.~\ref{eq:predict_n} has been modified to accommodate the combination of 
p$_0$ and p$_1$. Additionally, the measurements of 
$^{12}$C(${}^{12}$C,p$_i$)$^{23}$Na by Fang et al. \cite{Fang2013} 
performed at Notre Dame in the energy range 
3$\,<\,$${\rm E}_{\mathrm{cm}}$$<\,$6~MeV have also been used to predict 
the ${}^{12}$C(${}^{12}$C,n)$^{23}$Mg cross section~\cite{bucherNN2012}.  
In this case, up to $N$=6 possible decay channels are required for 
the prediction
calculated in Eq.~\ref{eq:predict_n}. 

\begin{figure}
    \includegraphics[width=20pc,trim=23mm 93mm 22mm 93mm,clip]{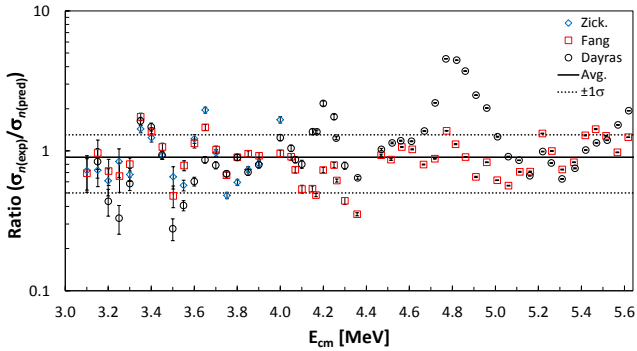}
    \caption{(Color online) The ratio of our 
$^{12}$C($^{12}$C,n)$^{23}$Mg cross section data $\sigma_{n(\text{exp})}$ 
to the two $\sigma_{n(\text{pred})}$ based on the 
$^{12}$C($^{12}$C,p)$^{23}$Na data from 
Zickefoose \cite{zickefoose} (blue diamonds) and from Fang et al. 
\cite{Fang2013} (red squares).
The solid black line shows the average ratio of the Zickefoose data 
below 4.0 MeV
while the dashed lines represent 1 standard deviation. As a comparison, 
the ratios of our $\sigma_{n(\text{exp})}$ to the Dayras prediction are 
shown as black circles. 
}
    \label{fig:predRatio}
\end{figure}

Figure~\ref{fig:predRatio} shows the ratios between our measured 
$^{12}$C($^{12}$C,n)$^{23}$Mg cross section $\sigma_{n(\text{exp})}$ and 
the two $\sigma_{n(\text{pred})}$ based on the Zickefoose and Fang 
proton data sets plotted as a function of E$_{\rm cm}$.  The average 
ratios (standard deviations) for the Zickefoose- and Fang-based 
predictions are 0.9(4) and 0.9(3), respectively.  The ratios to the 
Dayras calculation are also shown for comparison.  The large deviation at 
E$_{\rm cm}$$\simeq$\,4.8 MeV has been eliminated by our approach. 
The fluctuations, which are larger than the quoted statistical 
uncertainties, reflect the systematic errors associated with 
our extrapolation. They consist of the systematic errors in the 
proton measurements, the assumed entrance channel spin populations, and 
the TALYS calculation used in the prediction of 
$\sigma_{\mathrm n_i(\text{th})}/\sigma_{\mathrm p_i(\text{th})}$.
  To provide better consistency with the experimental 
$^{12}$C($^{12}$C,n)$^{23}$Mg data, our extrapolation has been 
renormalized by the factor 0.9.  We have adopted 0.4 as the systematic 
error in accordance with the Zickefoose-based prediction since that data 
set was used for the extrapolation, being the only one to reach 
sufficiently low energies.

The new cross section defined by our extrapolation and experimental data 
has been used to calculate the ${}^{12}$C(${}^{12}$C,n)$^{23}$Mg 
reaction rate by the following equation: 
\begin {equation} \label{eq:reaction_rate}
\langle\sigma v\rangle= \Big(\frac{8}{\pi \mu}\Big)^{1/2} 
\frac{1}{{kT}^{3/2}} \int_{E_{\mathrm{th}}}^{ \infty } 
\sigma(E)E\exp\Big(-\frac{E}{kT}\Big)\mathrm dE
\end {equation}

\begin{figure}
    \includegraphics[width=20pc,trim=0mm 0mm 0mm 58mm,clip]{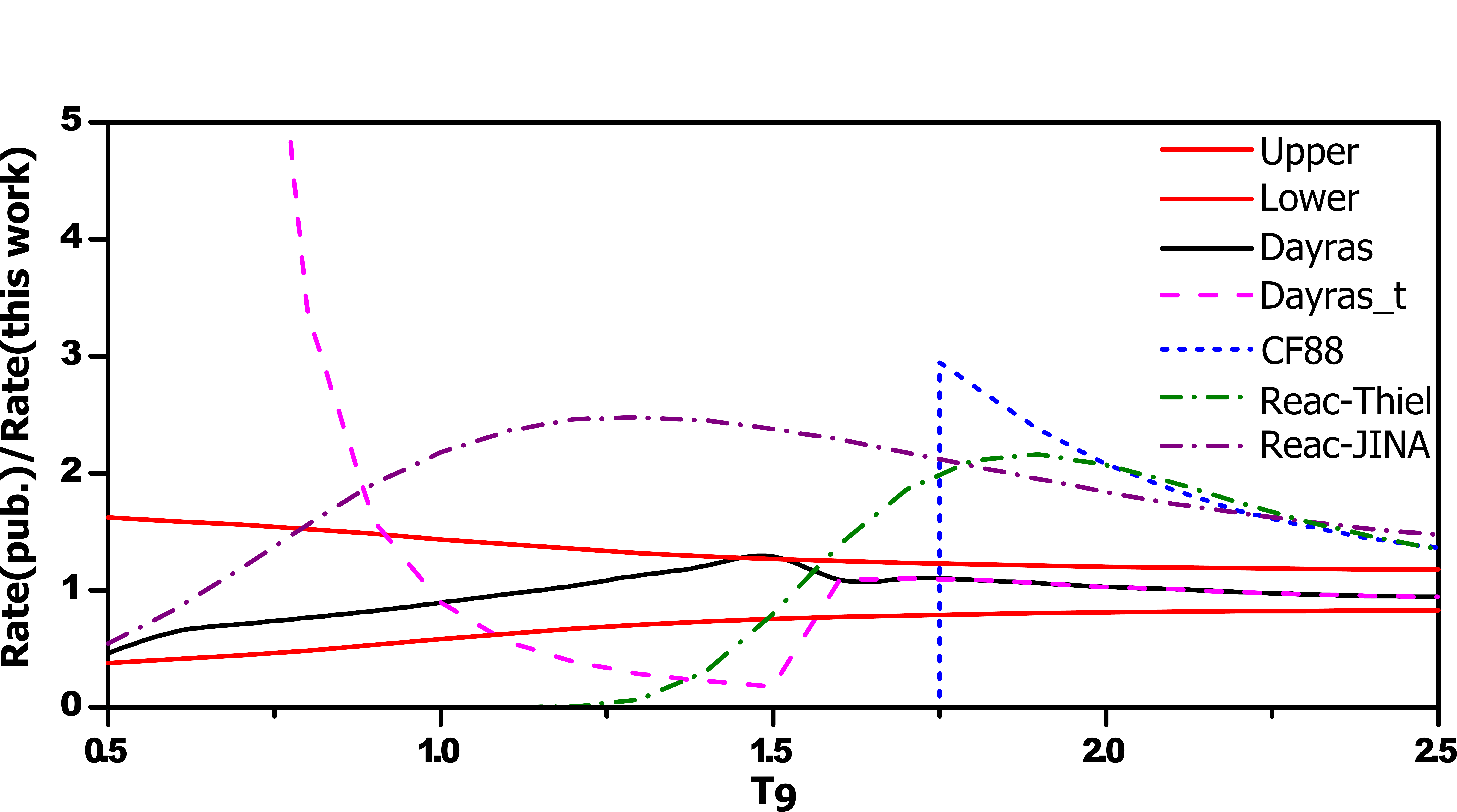}
    \caption{(Color online) The ratios of various published 
$^{12}$C($^{12}$C,n)$^{23}$Mg rates to the one 
determined in this work. The uncertainties of our new rate are indicated 
by the red lines.  The rate determined by Dayras is in good agreement  
with our new rate. The erroneous Dayras rate, stemming from a typo in 
the publication, is labeled Dayras\_t.  For comparison, we also show 
the rates from CF88 \cite{Caughlan1988} and two separate editions of 
REACLIB (from Thielemann et al. \cite{reaclib} and JINA \cite{jina}).}
    \label{fig:ccn_rate_comparison}
\end{figure}

To highlight the important stellar energy range for a typical carbon 
shell burning temperature $T_9$=\,1.1, the integrand of 
Eq.~\ref{eq:reaction_rate} (Gamow yield) is computed and shown in 
Fig.~\ref{fig:sFactor}. Our measurement covers about half of the 
stellar energy range. It reduces the dependence on extrapolation in 
the astrophysical reaction rate and provides a base for examining 
the systematic uncertainty of extrapolation.

Because of its endothermic character, a good fit of the 
${}^{12}$C(${}^{12}$C,n)$^{23}$Mg reaction rate was difficult to 
achieve using the standard REACLIB format. Following the convention of Dayras 
\cite{Dayras1977}, our ${}^{12}$C(${}^{12}$C,n)$^{23}$Mg rate has 
been normalized to the standard  CF88 ${}^{12}$C+${}^{12}$C \emph{total} 
fusion rate. The neutron branching ratio, $\beta_n$, has been fitted 
and listed below.
 \begin{eqnarray*} \label{eq:braching_ratio}
 \beta_{n}&=& 0.11954\,\exp\bigg[-\Big(\frac{0.16446}{{T_9}^3}
                +\frac{2.57495}{{T_9}^2}+\frac{1.94145}{T_9}\Big)\bigg] \\
                & &  (T_{9}\leq 1.5)  \\
 	        &=& 0.2212\,\big[1- \exp(-0.13597\;T_9+0.158)\big] \\
                & &  (1.5\leq T_{9} \leq 2.5)  \\
               &=& 0.048811\,\big[1 - \exp(-2.1124\; T_9+3.8791)\big]\\
               & &   (2.5\leq T_{9} \leq 5.0) \\
               &=& 0.04875  \quad  (T_{9} > 5.0)
 \end{eqnarray*}

The uncertainty for the reaction rate is estimated based on the error bars 
of experimental and extrapolated cross sections. 
Comparisons among the existing reaction rates are shown in 
Fig.~\ref{fig:ccn_rate_comparison}.
The various rates have been plotted as a ratio to our rate 
in order to compare 
them on a linear scale over a large temperature range.
It is seen that only the Dayras rate agrees with our new rate within 
the quoted uncertainty.  
At typical carbon shell burning temperatures $T_9$$\simeq$\,1.1--1.3, 
the uncertainty is less than 40\% which is sufficient for studying the weak 
s-process. The uncertainty is reduced to 20\% at 
$T_9$$\simeq$\,1.9--2.1 which is relevant for explosive carbon burning.

\begin{figure}
    \includegraphics[width=20pc]{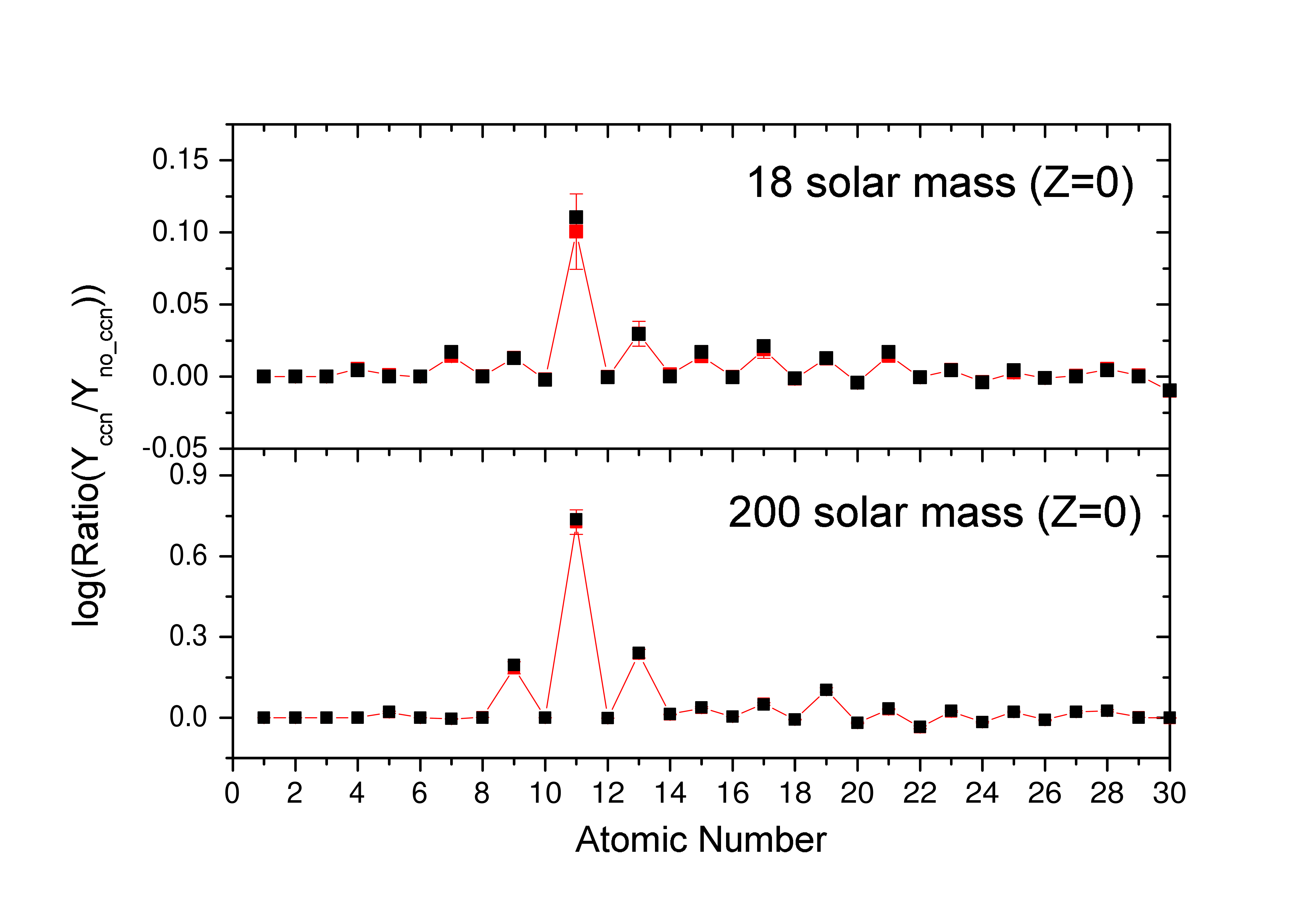}
    \caption{(Color online) The logarithmic ratio of elemental yields with 
${}^{12}$C(${}^{12}$C,n)$^{23}$Mg to those without 
${}^{12}$C(${}^{12}$C,n)$^{23}$Mg for 18\,M$_\odot$ (top) and 
200\,M$_\odot$ (bottom) Pop-III stars. The red points and their error 
bars correspond to the result obtained with the average and 
upper/lower limits determined by this work.The result obtained with 
the Dayras rate is shown as black points.}
    \label{fig:ccn_abund_ratio_pop3}
\end{figure}

The impact of ${}^{12}$C(${}^{12}$C,n)$^{23}$Mg on the 
nucleosynthetic pattern of a 200\,M$_\odot$ Pair Instability Supernovae 
(PI SNe) has been investigated using the 1D stellar evolution code, 
KEPLER \cite{Kepler1,Kepler2}. The ratio of the production yields with 
and without ${}^{12}$C(${}^{12}$C,n)$^{23}$Mg is shown in 
Fig.~\ref{fig:ccn_abund_ratio_pop3}. It is clear that this reaction 
is important for the nucleosynthesis of odd-Z elements such as F, Na, and 
Al. By including our ${}^{12}$C(${}^{12}$C,n)$^{23}$Mg rate in 
the calculation, the production of ${}^{23}$Na is increased by a factor of 
5 (0.7 dex) with an uncertainty less than $\sim$10\%. The yield of 
${}^{27}$Al is increased by nearly a factor of 2 (0.3 dex). 
We have also explored the impact of ${}^{12}$C(${}^{12}$C,n)$^{23}$Mg on 
an 18 M$_\odot$ Pop-III star. A moderate enhancement of up to 30\% is 
found for odd-Z elements.

The role of ${}^{12}$C(${}^{12}$C,n)$^{23}$Mg in Pop-I stars has 
been overlooked in most studies of the weak s-process because both CF88 
and REACLIB (from Thielemann \cite{reaclib}) essentially 
turn it off at carbon shell burning temperatures. 
To illustrate its impact on the nucleosynthesis in massive stars, an 
18\,M$_\odot$ Pop-I star has been investigated using KEPLER with 
two different scenarios: including and excluding 
${}^{12}$C(${}^{12}$C,n)$^{23}$Mg.
 By comparing these two production yields, an enhancement of $\simeq$10\% 
is found for a number of weak s-process isotopes, such as  
${}^{70}$Zn, ${}^{76}$Ge, ${}^{82}$Se, ${}^{86}$Kr, ${}^{85,87}$Rb and 
${}^{96}$Zr,  arising from the additional neutron production from 
${}^{12}$C(${}^{12}$C,n)$^{23}$Mg. 
The origins of these elements are rather complicated including He-, C-, 
Ne-burning, s-process in AGB stars and the r-process. Even within the
weak s-process there are a number of uncertainties that can affect the 
final abundance pattern 
\cite{Prantzos1990,Raiteri1991,Pignatari2010,Pignatari2013, Best2013}.  
Our result clarifies the ambiguities associated with the 
${}^{12}$C(${}^{12}$C,n)$^{23}$Mg rate. Furthermore, about 10\% 
enhancements are also observed for ${}^{46}$Ca and ${}^{60}$Fe.  
The production mechanism of ${}^{46}$Ca is important for the understanding 
of the ${}^{48}$Ca$/{}^{46}$Ca anomaly in meteorites \cite{Sorlin1993}, while 
the production of ${}^{60}$Fe is an important topic in $\gamma$-ray 
astronomy \cite{Wang2007}.

In summary, we have measured the ${}^{12}$C(${}^{12}$C,n)${}^{23}$Mg 
cross section for the first time within the Gamow window for the 
stellar carbon burning processes. Our measurement covers half of 
the important energy range. For the lower unmeasured energies, we 
have developed a novel extrapolation method based on the 
${}^{12}$C(${}^{12}$C,p)${}^{23}$Na channel.  A new reaction rate has 
been determined with, for the first time, a quantified uncertainty 
that satisfies the precision required from astrophysics models.  
As a result, the ambiguity arising from the uncertain 
${}^{12}$C(${}^{12}$C,n)$^{23}$Mg reaction rate has been eliminated. With 
our new rate, we find that ${}^{12}$C(${}^{12}$C,n)$^{23}$Mg is crucial 
for constraining
the production of Na and Al in Pop-III Pair Instability Supernovae, and 
it plays a non-negligible role in the production of weak s-process 
elements as well as the production of the important galactic 
$\gamma$-emitter ${}^{60}$Fe.

\begin{acknowledgments}
The authors would like to thank H. Esbensen (ANL) for providing the
spin population of the ${}^{12}$C+${}^{12}$C entrance channel,
 F.~Montes (MSU) for supplying two of the $^3$He
counters used in the experiment, 
and F.~Strieder (Bochum) for his suggestion of using a HOPG target.  
This work was supported by the NSF
under Grants No.~PHY-0758100 and No.~PHY-0822648, the
National Natural Science Foundation of China under Grant No. 11021504,
11321064, 11475228 and 11490564, 100 talents Program of the Chinese
Academy of Sciences, the Joint Institute for Nuclear Astrophysics, 
and the University of Notre Dame.
B.B. acknowledges support for the preparation of this manuscript by
the U.S. Department of Energy and Lawrence Livermore National
Laboratory under contract DE-AC52-07NA27344.  M.P. acknowledges
support from the ``Lend\"{u}let-2014" Programme of the Hungarian Academy 
of Sciences (Hungary), from 
SNF (Switzerland) and from the NuGrid collaboration.  
A.H.\ was supported by a Future Fellowship by
the Australian Research Council (FT120100363).  S.W. was supported by NASA 
(NNX14AH34G) and the UC Office of the President (12-LF-237070).
\end{acknowledgments}


\bibliography{References}

\end{document}